\documentclass[twocolumn,twoside,slac_two]{revtex4}
\usepackage{graphicx}
\usepackage{fancyhdr}
\usepackage{subfigure}
\usepackage{mdwlist}
\usepackage{xspace}
\def\deg{$^\circ$}
\def\Xmax{$X_{\rm max}$\xspace}
\def\gcm2{g/cm$^2$}

\def\Cerenkov{\v{C}erenkov\xspace}

\def\cref#1{Chapt.\,\ref{#1}}
\def\Cref#1{Chapter~\ref{#1}}
\def\sref#1{Sect.\,\ref{#1}}

\def\fref#1{Fig.\,\ref{#1}}

\def\inst#1{$^{#1}$}

\begin{document}

\title{Measurement of the properites of cosmic rays with the LOFAR radio
       telescope}

\author{
J\"org R.\ H\"orandel\inst{1,2,*}
A.~Bonardi\inst{1},
S.~Buitink\inst{1,3},
A.~Corstanje\inst{1},
H.~Falcke\inst{1,2,4,5},
B.~Hare\inst{6},
P.~Mitra\inst{3},
K.~Mulrey\inst{3},
A.~Nelles\inst{1,2,8},
J.P.~Rachen\inst{1},
L.~Rossetto\inst{1},
P.~Schellart\inst{1,9},
O.~Scholten\inst{6,7},
S.~ter Veen\inst{4},
S.~Thoudam\inst{1,10},
T.N.G.~Trinh\inst{6}, and 
T.~Winchen\inst{3}
}
\affiliation{
$^1$Department of Astrophysics, IMAPP, Radboud University Nijmegen, P.O. Box
9010, 6500 GL Nijmegen, The Netherlands\\
$^2$NIKHEF, Science Park Amsterdam, Amsterdam 1098 XG, The Netherlands\\
$^3$Astrophysical Institute, Vrije Universiteit Brussel, Pleinlaan 2, Brussels
1050, Belgium\\
$^4$ASTRON, Postbus 2, Dwingeloo 7990 AA, The Netherlands\\
$^5$Max-Planck-Institut f\"ur Radio Astronomie, Bonn, Germany\\
$^6$KVI-CART, Groningen University, P.O. Box 72, Groningen 9700 AB, The
Netherlands\\
$^7$Interuniversity Institute for High Energies, Vrije Universiteit Brussel,
Pleinlaan 2, 1050 Brussels, Belgium\\
$^8$now at: Department of Physics and Astronomy, University of California
Irvine, Irvine, CA 92697-4575, USA\\
$^9$now at: Department of Astrophysical Sciences, Princeton University,
Princeton, NJ 08544, USA\\
$^{10}$now at: Department of Physics and Electrical Engineering
 Linneuniversitetet, 35195 V\"axj\"o, Sweden\\
$^*$http://particle.astro.ru.nl
}

\begin{abstract}
High-energy cosmic rays, impinging on the atmosphere of the Earth initiate
cascades of secondary particles, the extensive air showers. The electrons and
positrons in the air shower emit electromagnetic radiation. This emission is
detected with the LOFAR radio telescope in the frequency range from 10 to 240
MHz. The data are used to determine the properties of the incoming cosmic
rays. The radio technique is now routinely used to measure the arrival
direction, the energy, and the particle type (atomic mass) of cosmic rays in
the energy range from $10^{17}$ to $10^{18}$ eV.  This energy region is of
particular astrophysical interest, since in this regime a transition from a
Galactic to an extra-galactic origin of cosmic rays is expected.  
For illustration, the LOFAR results are used to set constraints
on models to describe the origin of high-energy cosmic rays.
\end{abstract}

\maketitle

\thispagestyle{fancy}

\section{Introduction} \label{intro}
Cosmic rays (ionized atomic nuclei) impinge on the Earth with (kinetic)
energies covering a wide range from MeV energies up to beyond $10^{20}$~eV.  At
energies below $\sim 100$~MeV they are accelerated in energetic outbursts of
the Sun. At higher energies, the are assumed to originate in our Milky Way,
being accelerated in Supernova remnants, e.g.\
\cite{hessrxj1713,voelkrxj1713}.  At energies exceeding $10^{18}$~eV it becomes
increasinlgy difficult to magnetically bind the particles to our Galaxy. Thus,
particles with energies above $\sim10^{18}$~eV are usually considered to be of
extra-galactic origin.  A transition from a Galactic to an extra-galactic
origin of cosmic rays is expected at energies around $10^{17}$ to $10^{18}$~eV
\cite{behreview,naganowatson}.

Understanding the origin of cosmic rays in the transition region
($10^{17}-10^{18}$~eV) necessitates a precise measurement of the properties of
cosmic rays, namely their arrival direction (on the sky), their (kinetic)
energy, and their particle type (atomic mass $A$). The radio detection of air
showers provides a new tool for such measurements.

The flux of cosmic rays is steeply falling, approximately following a power law
$\propto E^{-3}$. In our region of interest, cosmic rays are only measured
indirectly, using large ground-based detector installations.  High-energy
cosmic rays impinging on the atmosphere, initiate cascades of secondary
particles, the extensive air showers.  The challenge of the indirect
measurements is to derive the properties of the incoming cosmic rays from
air-shower observations.  Most challenging is the measurement of the particle
type, since the sensitivity of air shower measurements is only proportional to
$\ln A$.  Intrinsic shower fluctuations allow to divide the measured cosmic
rays in up to five mass groups for the best experiments \cite{ricap07}.

\begin{figure*}[t]
 \centering
 \includegraphics[width=\textwidth]{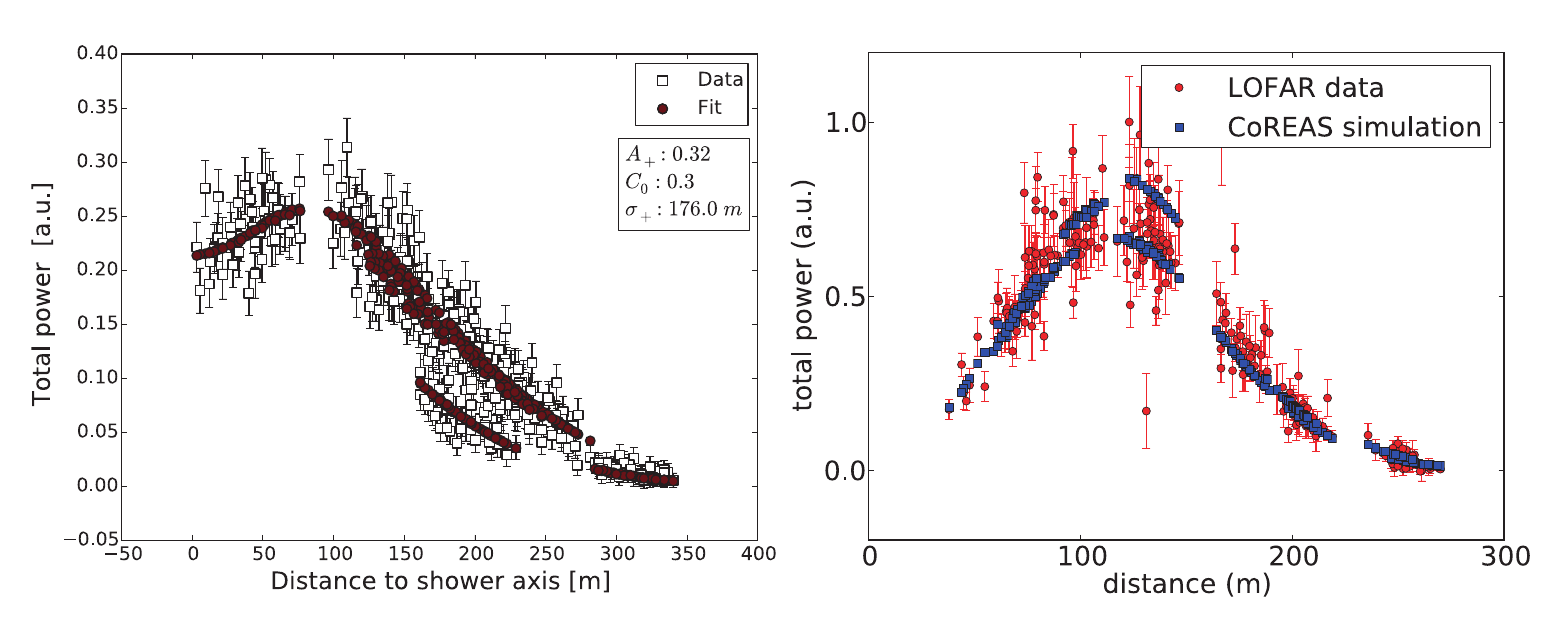}
 \caption{
Air showers measured with LOFAR \cite{Nelles:2014gma,Nelles:2014dja}. The total
power as a function of the distance to the shower axis in the frequency range
$30 - 80$ MHz (left) and $110 - 190$ MHz (right).
}
 \label{lat}
\end{figure*}

The radio measurement of air showers is briefly sketched in \sref{radio}.  The
method is used to determine the properties of cosmic rays as outlined in
\sref{properties}. One of the key results obtained is the mass composition of
cosmic rays in the transition region. Implications on our understanding of the
origin of cosmic rays will be discussed.

\section{Radio detection of air showers}\label{radio}
Many secondary particles in extensive air showers are electrons and
positrons. They emit radiation with frequencies of tens of MHz mainly
due to interaction with the magnetic field of the Earth.  Radio detection of
air showers is suitable to measure the properties of cosmic rays with nearly
100\% duty cycle \cite{Huege:2016veh,Schroder:2016hrv}.  

\subsection{LOFAR}
The LOFAR radio telescope \cite{vanHaarlem:2013dsa,Schellart:2013bba} is one of
the leading installations for the radio measurements of air showers.
LOFAR is a digital radio telescope. Its antennas are spread over several
European countries and are used together for interferometric radio observations
in the frequency range of $10-240$ MHz. The density of antennas increases
towards the center of LOFAR, which is located in the Netherlands.  Here, about
2400 antennas are clustered on an area of roughly 10 km$^2$ with increasing
antenna density towards the center. This high density of antennas makes LOFAR
the perfect tool to study features of the radio emission created by extensive
air showers. The radio antennas have been calibrated with in-situ measurements,
using a reference source and Galactic emission \cite{Nelles:2015gca}.

Air shower measurements are conducted based on a trigger received from an array
of scintillators (LORA) \cite{Thoudam:2014cfa,Thoudam:2015lba}, which results
in a read-out of the ring buffers that store the raw voltage traces per antenna
for up to 5 s.  LOFAR comprises two types of antennas. While air showers have
also been measured in the high frequency band ($110-240$ MHz), most air showers
are measured with the low-band antennas (LBA), which cover the frequency range
from $10-90$ MHz.  The LBAs are arranged in compact clusters of 96 antennas,
called stations.  Of every station either the inner group or the outer ring of
antennas (48 antennas each) can be used for cosmic-ray measurements at a given
time.

\subsection{Radio emission}
In the last years the radio technique has been established as a precise method
to measure the mass composition of cosmic rays.  In addition to LOFAR, radio
emission from extensive air showers has also been measured in detail by the
pioneering experiments LOPES \cite{lopesspie,radionature} and CODALEMA
\cite{codalema,Ravel:2003mi,Riviere:2009pr} and on larger scales by Tunka-REX
\cite{Schroder:2015dea} and the Auger Engineering Radio Array -- AERA at the
Pierre Auger Observatory \cite{Abreu:2011fb,Abreu:2012pi}.

The LOFAR measurements
together with the predictions of the CoREAS \cite{Huege:2013vt} simulation
package result in a complete understanding of the emission mechanisms. 
With
LOFAR the properties of the radio emission have been measured with high
accuracy \cite{jrh-icrc13,jrh-icrc15,Horandel:2015tpa} in the frequency range
$30-80$~MHz, which allows us to establish key features of the radio emission.

\subsection{Lateral distribution function of the radio signals}
The footprint of the
radio emission recorded at ground level is not rotationally symmetric, such as
e.g.\ the particle content of a shower 
\cite{Nelles:2014xaa,Nelles:2014gma}.
Radio emission is generated through
interactions with the Earth magnetic field, which yield a bean-shaped footprint
on the ground. The measured power is plotted as a function of the distance to
the shower axis in \fref{lat} in the frequency bands $30- 80$ MHz (left) and
$110-190$ MHz (right). For example at a distance of 200 m from the shower axis
in the left figure, ambiguities are visible in this one-dimensional projection:
the recorded signal strength is a function of the azimuth angle, which results
in the visible structure. The power density of the radio emission can be
parameterized by an analytical expression: a two-dimensional Gaussian function
is used to describe the approximately exponential fall-off at large distances
form the shower axis. A second (smaller) two-dimensional Gaussian function is
subtracted from the first one to describe the ring structure of the signal
close to the shower axis. To reproduce the observed bean shape, the centres of
both Gaussian functions are slightly offset.

\begin{figure}[t]
 \centering
 \includegraphics[width=\columnwidth]{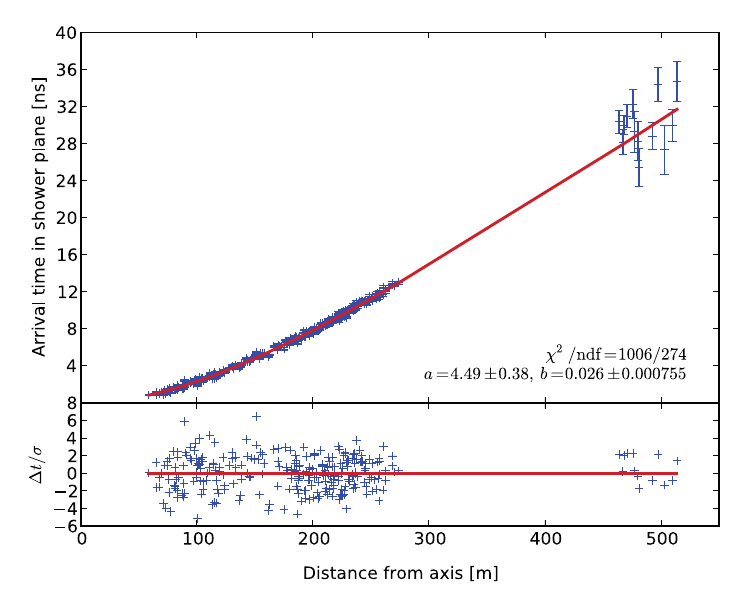}
 \caption{ Arrival time of the signals as a function of the distance to the
shower axis \cite{Corstanje:2014waa}. The lower graph illustrates the
arrival time differences with respect to a fit (hyperboloid).}
 \label{shape}
\end{figure}

\subsection{Shape of the shower front} 
The
precise shape of the radio wave front is a long-standing issue
\cite{Corstanje:2014waa,Corstanje:2016mia}. 
In the
literature different scenarios have been discussed: a spherical shape or a
conical shape. The LOFAR findings clearly indicate that a hyperboloid is the
best way to describe the shape of the measured wave front, confirming first
hints from LOPES \cite{Schroeder2011,Apel:2014usa}. A hyperboloid
asymptotically reaches a conical shape at large distances from the shower axis
and can be approximated as a sphere close to the shower axis. A measured wave
front of a shower registered with LOFAR is shown in \fref{shape}.  The time
difference relative to a plane is plotted as a function of a distance to the
shower axis. The line indicates a fit of a hyperboloid to the measured data.
The lower part of the graph shows the time differences of the individual
antennas with respect to the fit function.

\begin{figure}[t]
 \centering
 \includegraphics[width=\columnwidth]{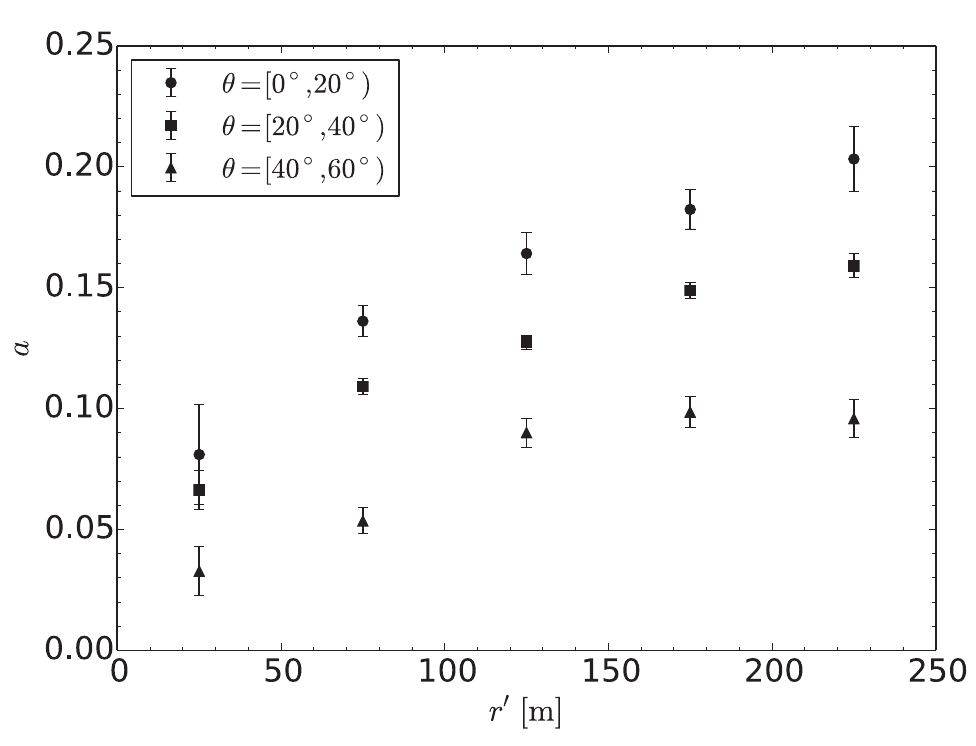}
 \caption{Radio emission processes in the atmosphere \cite{Schellart:2014oaa}.
The ratio between Askaryan effect and geomagnetic emission is plotted as a
function of the distance to the shower axis for showers with different zenith
angles, as indicated, and measured by LOFAR.}
 \label{a}
\end{figure}

\subsection{Polarization of the radio signal}
The radio emission in extensive air
showers originates from different processes  
\cite{Schellart:2014oaa,Scholten:2016gmj}. 
The dominant mechanism is of
geomagnetic origin
\cite{kahnlerche,allanrev,Falcke:2002tp,radionature,Huege:2012vk,Ardouin:2009zp}.
Electrons and positrons in the shower are accelerated in opposite directions by
the Lorentz force exerted by the magnetic field of the Earth. The generated
radio emission is linearly polarized in the direction of the Lorentz force $(v
\times B)$, where $v$ is the propagation velocity vector of the shower
(parallel to the shower axis) and $B$ represents the direction and strength of
the Earth magnetic field. A secondary contribution to the radio emission
results from the excess of electrons at the front of the shower (Askaryan
effect) \cite{askaryanexcess}. This excess is built up from electrons that are
knocked out of atmospheric molecules by interactions with shower particles and
by a net depletion of positrons due to annihilation. This charge excess
contribution is radially polarized, pointing towards the shower axis. The
resulting emission measured at the ground is the sum of both components.
Interference between these components may be constructive or destructive,
depending on the position of the observer/antenna relative to the shower. The
emission is strongly beamed in the forward direction due to the relativistic
velocities of the particles.  Additionally, the emission propagates through
the atmosphere, which has a non-unity index of refraction that changes with
height. This gives rise to relativistic time-compression effects, most
prominently resulting in a ring of amplified emission around the \Cerenkov
angle, see \fref{lat}, right. By precisely measuring the
polarization direction of the electric field in each LOFAR antenna we determine
the relative contribution of the main emission processes, thus clarifying the
emission processes in air showers. The parameter $a$ gives the ratio of the
charge excess contribution to the geomagnetic radiation. This ratio is depicted
in \fref{a} as a function of the distance to the shower axis. Measurements
for showers with different zenith angles are shown. The figure illustrates that
the contribution through the Askaryan effect increases with increasing distance
to the shower axis and it is more pronounced for vertical showers (with small
zenith angle). Corresponding investigations at AERA yield an average value
$a=14\%\pm2\%$ \cite{Aab:2014esa}.

\subsection{First quantitative measurements in the frequency range 120-240 MHz}
Radio emission from extensive air showers has also been
recorded with the high-band antennas in the 200~MHz frequency domain
\cite{Nelles:2014dja}. The
measured power is depicted in \fref{lat}, right as a function of the distance
to the shower axis. A clear maximum is visible at distances around 120~m,
indicating a clear (\Cerenkov) ring structure. Such rings are predicted from
theory: relativistic time compression effects lead to a ring of amplified
emission, which starts to dominate the emission pattern for frequencies above
$\sim 100$~MHz. The LOFAR data clearly confirm the importance to include the
index of refraction of air as a function of height into calculations of the
radio emission (see also \cite{Corstanje:2017djm}).

\begin{figure}[t]
 \centering
 \includegraphics[width=\columnwidth]{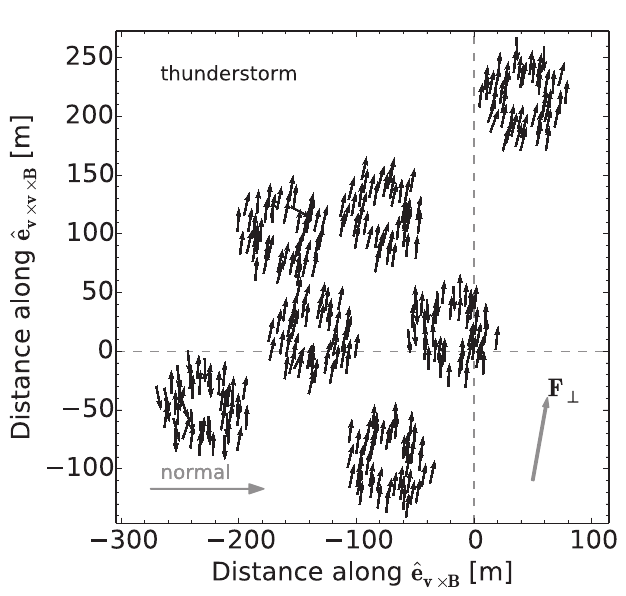}
 \caption{Footprint of the radio emission of an air shower, which developed
during a thunderstorm \cite{Schellart:2015kga}.}
 \label{thunder}
\end{figure}

\subsection{Probing atmospheric electric fields during thunderstorms}
Radio detection of air showers is also
used for auxiliary science, such as the measurements of electric fields in the
atmosphere during thunderstorms 
\cite{Schellart:2015kga,Trinh:2015nuw}. 
The footprint of the radio emission from an
air shower, which developed during a thunderstorm is shown in \fref{thunder}.
The intensity and polarization patterns of such air showers are radically
different from those measured during fair-weather conditions. The figure
illustrates the polarization as measured with individual LOFAR antennas
(arrows) in the shower plane. LOFAR antennas are grouped into circular
stations, of which seven are depicted. An arrow labelled “normal” indicates the
expected polarization direction for fair weather conditions. The position of
the shower axis, orthogonal to the shower plane, is indicated by the
intersection of the dashed lines. With the use of a simple two-layer model for
the atmospheric electric field, these patterns can be well reproduced by
state-of-the-art simulation codes. This in turn provides a novel way to study
atmospheric electric fields.\\[\baselineskip]

\begin{figure}[t]
 \centering
 \includegraphics[width=\columnwidth]{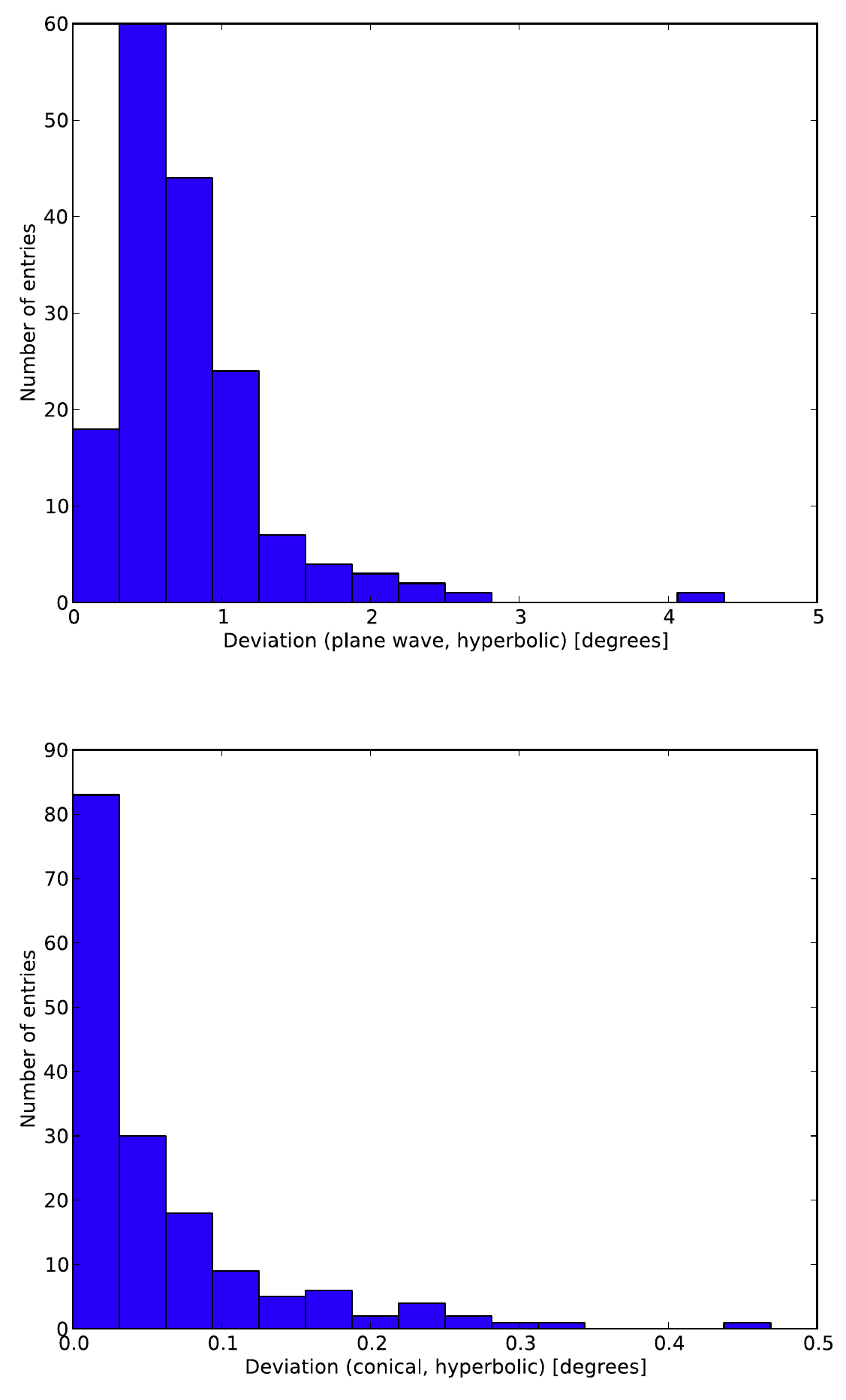}
 \caption{
Angular difference between reconstructed shower axis direction for three
wavefront shape assumptions. Assuming a planar wavefront shape typically
introduces an error in the direction of up to $\sim 0.1$\deg, when the shape is
in fact hyperbolic (top plot). The differences in reconstructed direction
between a conical and hyperbolic wavefront shape are approximately a factor of
ten smaller (bottom plot) \cite{Corstanje:2014waa}.
}
 \label{pointing}
\end{figure}

These measurements, together with findings from other groups
\cite{Revenu:2009si,Marin:2011bga,Belletoile:2015rea,Ardouin:2009zp,Aab:2014esa}
help to understand the emission processes in the atmosphere and to quantify the
contributions of the two mechanisms, being responsible for the radio emission
of air showers -- namely the geomagnetic effect (i.e.\ charge separation in the
geomagnetic field) and the Askaryan effect (charge excess in the shower front).
This allows to reconstruct the properties of the incoming cosmic rays from the
radio measurements as explained in the following section.

\section{Properties of cosmic rays}\label{properties}
Ultimate goal is to derive the properties of the incoming cosmic ray from the
radio measurements, namely their arrival direction (via the precise measurement
of the arrival time of the radio waves in the shower front, as described
above), their energy, and their particle type/atomic mass. The parameters of
the function to model the intensity pattern of the radiation on the ground (as
described above) are sensitive to the properties of the shower-inducing cosmic
rays. The integral of the measured power density is proportional to the shower
energy. The width of the measured footprint is proportional to the distance
from the antennas to the position of the shower maximum, which in turn can be
used to determine \Xmax and the mass of the cosmic ray.

\subsection{Direction}
The excellent time resolution of LOFAR with ns accuracy allows to measure the
shape of the shower front \cite{Corstanje:2014waa}.  In order to estimate the
accuracy of the measurement of the arrival direction of the shower, the same
measured air showers have been reconstructed with different assumptions (plane,
sphere, hyperboloid) as depicted in \fref{pointing}. These investigations
indicate an uncertainty for the direction reconstruction of the order of
0.1\deg to 0.5\deg.
 
\begin{figure}[t]
 \centering
 \includegraphics[width=\columnwidth]{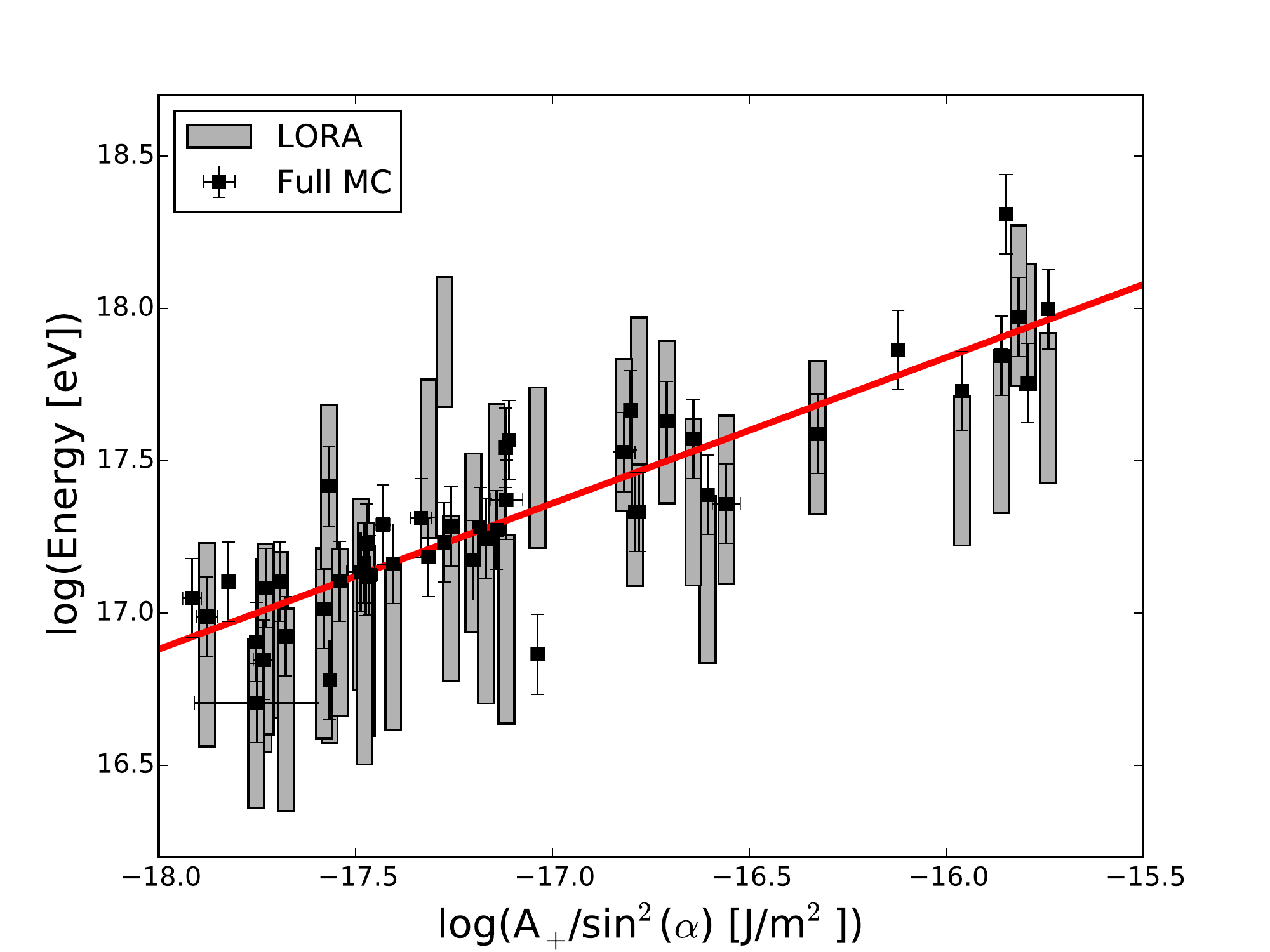}
 \caption{The energy as obtained from the particle detectors LORA
(grey bands) and the
energy from the full Monte-Carlo approach (black squares) are shown as a
function of the fit parameter $A_+$, i.e.\ the normalization of the
two-dimensional Gaussian function. Air showers with no grey bands are more
horizontal than 45\deg.  The straight line shows the best fit to the full
Monte-Carlo energies. Updated from \cite{Nelles:2014gma}.
}
 \label{energy}
\end{figure}

\subsection{Energy}
The recording of radio signals in the frequency range of interest ($30-80$ MHz)
provides an excellent calorimetric measure of the energy contained in the
electromagnetic component of the air shower and, thus, provides a good measure
of the energy of the shower-inducing particle.  The integral over the measured
energy fluence distribution on the ground is proportional to the energy of the
incoming cosmic ray.  
This is illustrated in \fref{energy}, the shower energy is obtained in two ways: measured with the LORA scintillator array and obtained from Monte-Carlo simulations. The results of both methods are plotted against the overall normalization parameter $A_+$ of the two-dimensional Gaussian function,
for details see \cite{Nelles:2014gma}.
A resolution around 30\% for the cosmic-ray energy is
obtained with LOFAR \cite{Buitink:2014eqa}.
Similar investigations at the Pierre Auger Observatory indicate that a
resolution around 25\% is possible for high-quality showers
\cite{Aab:2015vta,Aab:2016eeq}.

\begin{figure}[t]
 \centering
 \includegraphics[width=\columnwidth]{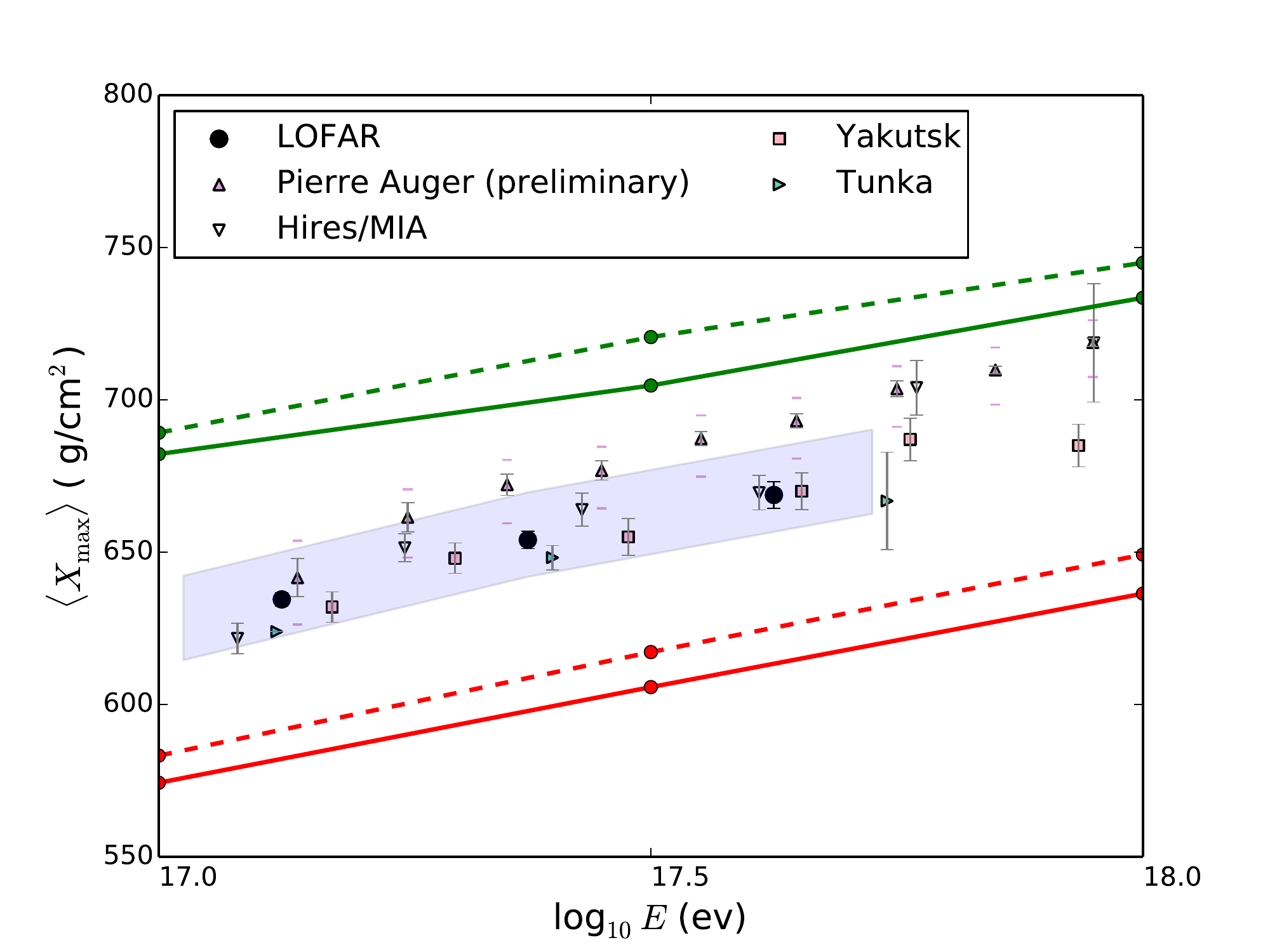}
 \caption{Average depth of the shwoer maximum \Xmax as a function of cosmic-ray
energy  \cite{Buitink:2016nkf}. The LOFAR radio results are compared to optical
measurements. For details and references see \cite{Buitink:2016nkf}.
The lines represent predictions for protons (green) and iron nuclei (red)
for the hadronic interaction models QGSJETII.04 (solid) and EPOS-LHC (dashed).
}
 \label{xmax}
\end{figure}

\begin{figure*}[t]
 \centering
 \includegraphics[width=0.99\textwidth]{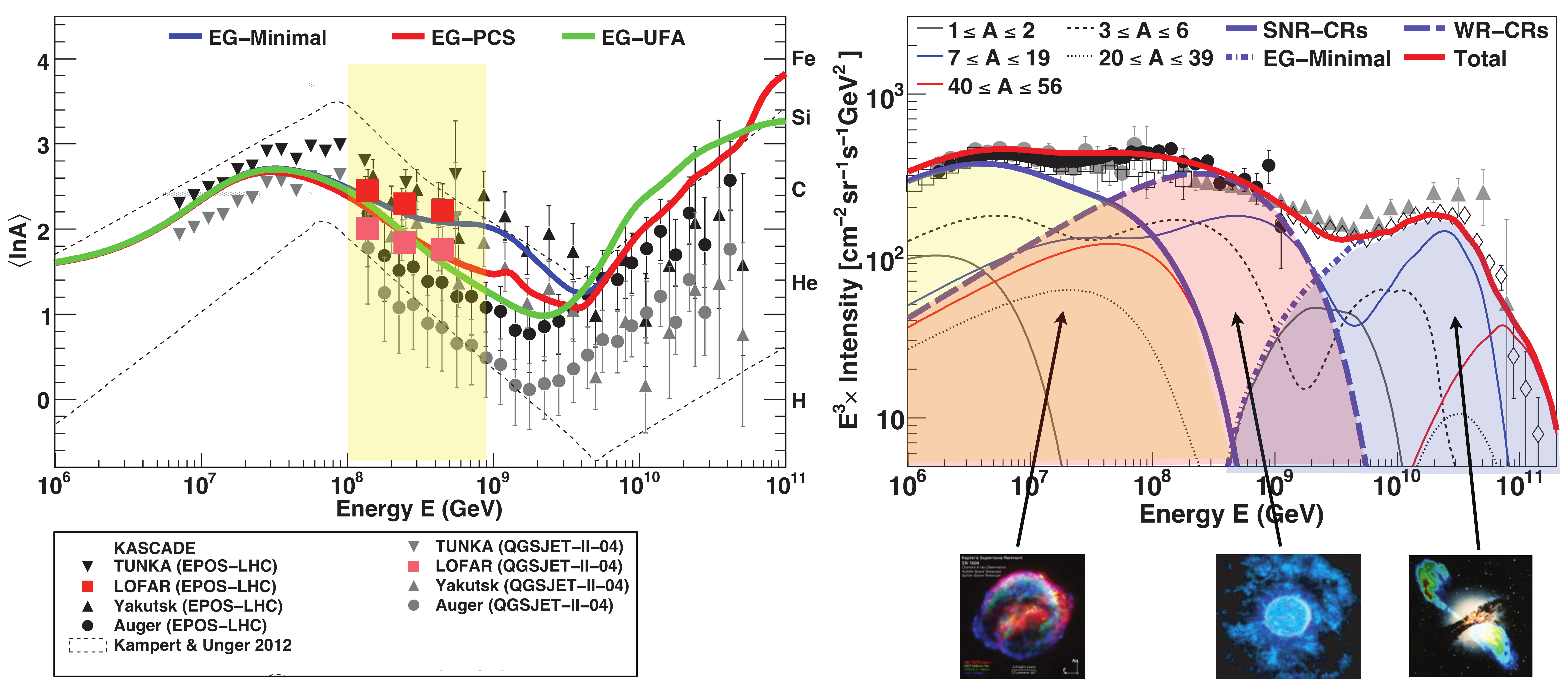}
 \caption{Left: Mean logarithmic mass of cosmic rays as a function of energy,
    for details, see \cite{Thoudam:2016syr}.
 Right: Three-component model of the origin of cosmic rays according to
   \cite{Thoudam:2016syr}: 'regular' supernovae, Wolf Rayet component, and
  an extra-galactic component.}
 \label{lna}
\end{figure*}

\subsection{Particle type}
The good agreement between the measurements and the predictions of the CoREAS
code is  essential to identify the type of incoming cosmic ray.  This is
inferred from the (atmospheric) depth of the shower maximum \Xmax, one of the
standard measures to estimate $\ln A$.  
Radio measurements of air showers are used to derive \Xmax
\cite{Apel:2014jol,Kostunin:2013iaa,Glaser:2016txi}.

To measure \Xmax with LOFAR \cite{Buitink:2014eqa,Buitink:2016nkf} we analyse
simultaneously measurements of the radio emission and the particle detectors.
The arrival direction and energy of each cosmic ray are determined first. Then,
simulations for primary protons and iron nuclei are conducted for each measured
shower with its corresponding direction and energy.  Due to the intrinsic
shower fluctuations is it sufficient to simulate only protons and iron nuclei
to cover the parameter space in \Xmax. The predictions for the signals in the
particle detectors and the radio antennas are compared on a statistical basis
to the measured values. This method is used to determine \Xmax with an accuracy
of better than $\sim 20$~\gcm2 with the dense LOFAR core, thus, reaching the
state of the art -- the uncertainty of the Pierre Auger Observatory
fluorescence detector.  The \Xmax values obtained are depicted as a function of
energy in \fref{xmax} together with other measurements. The latter apply
different techniques, namely measuring \Cerenkov and fluorescence light from
the air showers. The figure illustrates the good agreement between the radio
measurements and the established optical methods.

An important systematic effect is the refractive index of the air $n$  and its
dependence on atmospheric conditions, such as pressure, temperature, and
humidity. Recent investigations show that the refractivity $N=10^6 (n-1)$ can
have relative variations of the order of 10\% depending on atmospheric
parameters, typical variations are of the order of 4\%. They result in a
systematic uncertainty of the depth of the shower maximumof a few ($3.5-11)$
\gcm2 \cite{Corstanje:2017djm}.

The measured values for the depth of the
shower maximum \Xmax are converted to the mean logarithmic mass of cosmic
rays
$$\langle\ln
A\rangle=\left(\frac{X_{\rm max}-X_{\rm max}^{\rm p}}
    {X_{\rm max}^{\rm Fe}-X_{\rm max}^{\rm p}}\right)\times\ln
A_{\rm Fe} .$$
This necessiates predictions for the depth of the shower maximum for impinging
protons and iron nuclei, $X_{\rm max}^{\rm p}$ and $X_{\rm max}^{\rm Fe}$, 
respectively.  
These are illustrated as lines in \fref{xmax}: protons (green) and iron nuclei (red).
The resulting mean mass is depicted in \fref{lna} (left) as a
function of energy for the LOFAR results together with the world data set
\cite{Thoudam:2016syr}.  Two hadronic interaction models are used (EPOS-LHC and
QGSJETII.04, dashed and solid lines in \fref{xmax}, respectively) to interpret
the data. Both interaction models are tuned to LHC data but uncertainties
remain, when extrapolating to the cosmic-ray parameter space.

\subsection{Origin of cosmic rays}
To understand the implications of the LOFAR measurements and the available
world data set from direct and indirect measurements a model has been developed
to consistently describe the observed energy spectrum and mass composition of
cosmic rays with energies up to about $10^{18}$~eV \cite{Thoudam:2016syr}.
We assume that the bulk of Galactic cosmic rays is accelerated by strong
Supernova remnant shock waves \cite{Thoudam:2014sta}.
Our study shows that a single Galactic component with rigidity-dependent energy
cut-offs in the individual spectra of different elements cannot explain the
observed all-particle spectrum at energies exceeding $\sim2\cdot10^{16}$~eV.
Similar findings have already been obtained earlier \cite{Hillas:2005cs}.
We discuss two approaches for a second component of Galactic cosmic rays:
re-acceleration at a Galactic wind termination shock and Supernova explosions
of Wolf-Rayet stars.
The latter scenario can explain almost all observed features in the
all-particle spectrum and the mass composition of cosmic rays up to $\sim
10^{18}$~eV, when combined with a canonical extra-galactic spectrum as expected
from strong radio galaxies or a source population with similar cosmological
evolution. The resulting spectrum is shown in \fref{lna} (right).  In this
two-component Galactic cosmic-ray model, the 'knee' at $\sim4\cdot10^{15}$~eV
and the 'second knee' at $\sim 4\cdot10^{17}$~eV in the all-particle spectrum
are due to the fall-offs of the first and second Galactic cosmic-ray components,
respectively.

\section{Conclusions}
The radio detection of extensive air showers enables us to measure the
properties of cosmic rays above energies exceeding $10^{17}$~eV with high
precision of $\sim 0.1^\circ - 0.5$\deg for the arrival direction, $\sim30\%$
for the energy, and to better than $\sim20$~g/cm$^2$ for the depth of the
shower maximum \Xmax.

To illustrate the potential of the LOFAR radio measurements
we developed a model to consistently describe the observed energy spectrum and
mass composition of cosmic rays from GeV energies up to $10^{20}$~eV.
We adopt a three component model: 'regular' cosmic rays being accelerated in
Supernova remnants up to $\sim10ß^{17}$~eV, a second Galactic component,
dominating the all-particle flux between $\sim10^{17}$ and $\sim10^{18}$~eV
from cosmic rays being accelerated by exploding Wolf-Rayet stars, yielding a
strong contribution of He and CNO elements, and, finally, an extra-galactic
contribution at energies above $\sim10^{18}$~eV.


\section{Acknowledgements}
JRH is grateful to the organizers of the ECRS for their kind invitation.



\end{document}